\documentclass[a4paper,11pt]{article}
\evensidemargin=\oddsidemargin
\usepackage{bm}
\usepackage{float}
\usepackage{latexsym}
\usepackage{dcolumn}
\usepackage{amsfonts,amssymb}
\usepackage{graphicx}
\usepackage{epsfig}
\usepackage{psfrag}
\usepackage{nccmath}
\usepackage{moresize}
\usepackage{enumerate}
\usepackage[hang,flushmargin]{footmisc}
\usepackage[titletoc,toc]{appendix}
\usepackage{lipsum}
\usepackage{subcaption}
\usepackage{hyperref}
\usepackage{rotating}
\usepackage{multirow}
\usepackage{multicol}
\usepackage{xfrac}
\usepackage{mathtools}
\usepackage{nccmath}
\usepackage{cite}
\usepackage{placeins}
\usepackage{tikz}
\usetikzlibrary{positioning}

\graphicspath{{figures/}} 
\usepackage[font=small,labelfont=bf]{caption}
\usepackage{wrapfig}
\usepackage{authblk}
\usepackage{cite}
\usepackage[utf8]{inputenc}
\usepackage[T1]{fontenc}
\usepackage{enumitem}
\usepackage{blindtext}

\newcommand{\be}{\begin{equation}}
\newcommand{\ee}{\end{equation}}
\newcommand{\bea}{\begin{eqnarray}}
\newcommand{\eea}{\end{eqnarray}}

\title{  
  \bf Composite pseudo Nambu Goldstone Quintessence}

\author{Mayukh~R.~Gangopadhyay\thanks{mayukhraj@gmail.com}}
\affil[1]{\small \it Centre For Cosmology and Science Popularization, SGT University, Gurugram, Haryana-122505, India.}
\author[1]{Nilanjana Kumar\thanks{nilanjana.kumar@gmail.com}}
\author[2]{{Ankan Mukherjee}\thanks{ankan.ju@gmail.com}}
\affil{\small \it Department of Physics, Bangabasi College, Kolkata 700009, India.}
\author[1]{Mohit K. Sharma\thanks{mr.mohit254@gmail.com}}
\begin{document}
\date{}
\maketitle
\begin{abstract}
A pseudo-Nambu Goldstone Boson (pNGB) arising from  the breaking of a global symmetry ($G\rightarrow H$) can be one of the most promising candidate for the quintessence model, to explain the late time acceleration of our universe. Motivated from the Composite Higgs scenario, we have investigated the case where the pNGB associated with $SO(N)/ SO(N-1)$ develops a potential through its couplings with the particles that do not form the complete representations of $G$.
The Coleman Weinberg (CW) potential is generated via the external particles in the loop which are linked with the strongly interacting dynamics. 
This model of Dark Energy (DE) is tested against several latest cosmological observations such as supernovae data of Pantheon, Baryon Acoustic Oscillation (BAO), Redshift-space distortion (RSD) data etc.
We have shown that the model predicts cosmological parameters well within the allowed range of the observation and thus gives a well motivated model of quintessence. We have found that the fit against all data prefers sub-Planckian value of the pNGB field decay constant. Moreover, we have shown distinctively that the cosmological observations limit the value of the model parameter in such a way that the strong sector dynamics is highly constrained. 
\end{abstract}

\maketitle
\vspace{0.0001in}
\tableofcontents
\baselineskip=15.0pt
\section{Introduction}
\setcounter{equation}{0}
\setcounter{figure}{0}
\setcounter{table}{0}
\setcounter{footnote}{1}
\label{sec:intro}

Recent high-precision cosmological observations have indicated that the universe is currently undergoing a phase of acceleration \cite{Riess:1998cb,Perlmutter:1998np,planck18,PlanckXX,Planck:2018vyg,pantheon}. These observations also suggest that the universe may have undergone a period of accelerated expansion known as inflation during its early stages \cite{Liddle,cmbinflate}. While the scalar field is the primary candidate for explaining early universe expansion, In the case of late-time acceleration, there are various candidates, including scalar fields. This phenomenon is generally attributed to the dominance of exotic matter's energy density, which has negative pressure, over the total energy budget of the universe. This behavior is commonly referred to as Dark Energy (DE) dynamics \cite{CRS-quin,Carroll:2000fy,Peebles:2002gy,Padmanabhan:2002ji,tsuj-quin,CST-rev,FTH-rev,AT-book}.


The premise of the General Theory of Relativity (GR) suggests that the universe is undergoing an accelerated expansion, which has been observed from various observations. However, one can propose modified theories of gravity that go beyond GR to account for this expansion, provided that they pass the observational constraints\cite{tsuji}. Currently, the cosmological constant $\Lambda$ is the leading candidate to explain the existence of Dark Energy (DE) both observationally and from a simplistic point of view. Introducing a single parameter $\Lambda$ in the Einstein equation can fit all observations very accurately. However, the extremely small value of $\Lambda^{1/4}(\approx 2\times 10^{-3}eV)$, which is intermediate between particle physics and cosmology, presents a problem from the perspective of quantum field theory \cite{Padmanabhan:2002ji}. This tiny value is hard to reconcile with our present understanding of theoretical particle physics.


The concept of Dark Energy (DE) and the accelerated expansion of the universe can also be understood through the idea of scalar field dynamics within the realm of GR, known as ``quintessence'' in the context of DE \cite{CST-rev,tsuj-quin,ratra1988,Zlatev:1998tr,Sahni:1999qe,Amendola:1999er,Sahni:2002kh}. This idea is more suitable for meeting the theoretical demands due to the field's dynamical evolution. Recent studies of quintessence scalar field models for both inflation and late-time cosmic acceleration can be found in \cite{Geng:2015fla,Hossain:2018pnf,Durrive:2018quo,Sangwan:2017kxi,Ahmad:2019jbm,Rajvanshi:2019wmw,Elizalde:2022rss,Frasca:2022vvp}. While the concept of a scalar field is supported by theoretical particle physics, the ultra-light nature of the field required to produce observable effects makes it challenging to develop a model that maintains the flatness of the potential from quantum corrections.
The recently observed tension between the early time and late time measurement of the Hubble expansion rate has been studied in \cite{mohit}-\cite{mohit2} in the context of dark matter dark energy interaction.\\*
Quintessence in the context of Dark Energy and the accelerated expansion of the universe is further strengthened by the possibility of a pseudo-Nambu-Goldstone Boson (pNGB) arising from the spontaneous breaking of global symmetry ($G\rightarrow H$) with a flat potential maintained due to the shift symmetry. This idea was proposed in the early stages of quintessence realization \cite{frieman} and later constrained from data \cite{liddle2015}, while the concept of strongly coupled quintessence was first introduced in \cite{damico}.\footnote{This mechanism can  also lead to various interesting dynamics of early time accleration \cite{freese,croon, mrg1,hussain1}.} This paper focuses on late-time acceleration due to a composite pNGB quintessence model inspired by Composite Higgs like scenarios, where the one-loop potential is generated via the Coleman Weinberg mechanism. This model could appear even near the Planck scale, unlike standard scenarios, where the compositeness scale is not large \cite{Rosenlyst:2021tdr}. This paper aims to investigate whether the pNGB quintessence follows the same type of potential as in Composite Higgs like models \cite{Contino:2010rs}.


In this paper, we present a new approach to the quintessence field in the context of Composite Models, where a global $SO(N)$ symmetry breaking to $SO(N-1)$ produces $N-1$ massless Goldstone fields, one of which acts as the quintessence field. The authors consider the minimal Composite Higgs Model based on $SO(5)/SO(4)$ \cite{Contino:2003ve} and a scalar field, which gives rise to a potential with additional oscillating terms, unlike the simplified shift symmetric potential. The study of parameters in this potential is not ad-hoc but well motivated from Composite Higgs like scenarios. The analysis focuses only on the theoretical implications of the quintessence dynamics, while taking into account observational constraints.


In the realm of pNGB quintessence, previous literature has suggested that achieving the desired range of the equation of state parameter ($w$) for a pNGB quintessence requires a super-Planckian breaking scale. However, this approach poses a challenge from the perspective of effective field theory (EFT), and the super-Planckian regime is known to be inherently unstable due to non-perturbative quantum gravity effects. In contrast, we propose a model based on the CW mechanism, where we aim to confirm whether a sub-Planckian value of the breaking scale ($S$) can indeed satisfy the desired range of $w$ and account for the dynamics of the pNGB quintessence, which are governed by the presence of two oscillating terms in the potential.

The paper is structured as follows. Section \ref{review} provides a brief overview of the mechanism behind the quintessence field production, inspired by Composite Higgs theories. Section \ref{implications} explores the implications of the proposed model in late-time cosmology, while the analysis of the model's consistency with observational results is presented in Section \ref{dataa}. The paper concludes with a summary of findings and discussions in Section \ref{discussions}.

\section{Goldstone Quintessence}
\label{review}
\setcounter{equation}{0}
\setcounter{figure}{0}
\setcounter{table}{0}
\setcounter{footnote}{1}
We consider a scenario where the quintessence is a 
pseudo Nambu Goldstone boson (pNGB) coming from the global symmetry 
breaking of $G\rightarrow H$, and $S$ is the spontaneous symmetry breaking 
scale. The pNGB does not acquire 
a tree level potential due to the shift symmetry.
But it is possible to generate a potential in the loop level via the Coleman Weinberg (CW) mechanism, when the external particles are the 
source of explicit symmetry breaking.  
The scale of explicit symmetry breaking is lower but close to $S$. 
For this purpose, we choose the 
breaking of $SO(N)/ SO(N-1)$, giving $N-1$ massless Goldstone fields. 
Particularly we have picked up the spinoral representation 
of this group to ensure the flatness 
of the potential for slow rolling 
quintessence\footnote{For detail we refer to \cite{croon}, 
where the same potential has been studied for the inflationary scenario.}.
The CW potential in this case takes the form,
\begin{equation}
V(\phi)=\alpha\cos{\left(\frac{\phi}{S}\right)} + \beta\sin^2{\left(\frac{\phi}{S}\right)}
\end{equation}
where, $\phi$ is the quintessence field, 
$\alpha$ and $\beta$ are the fermionic and gauge contributions in the CW potential, given by,\\
\begin{equation}
\alpha = 2 N_C \int \frac {d^4p_E}{(2\pi)^4}\Big(\frac{\Pi_1^R}{\Pi_0^R}-\frac{\Pi_1^L}{\Pi_0^L}\Big);~~
\beta = \int \frac {d^4p_E}{(2\pi)^4}\Big(\frac{3(N-2)}{4}\frac{\Pi_1^A}{\Pi_0^A}- 2 N_c \frac{M^2}{pE^2\Pi_0^L\Pi_0^R}\Big).
\end{equation}
$\Pi^A_{0,1}(p)$ are the scale dependent form factors, which represent the 
integrated-out dynamics of the strong sector.
Studies of Goldstone motivated Dark Energy models \cite{liddle,dutta,ruchika} have been 
done with the potential of the form $\sim (1+ \cos(\Phi/S))$. 
Whereas, in this paper we have two  oscillating terms with 
coefficients $\alpha$ and $\beta$, both playing an important role in the dark energy dynamics.

Moreover, the relation between $\alpha$ and $\beta$ can shed lights on the masses and decay constants of the particles that contribute in the loop. The form factors can be thought of as sum of towers of resonances. If we assume that the form factors can be well
approximated by considering only the contribution from the lightest of these resonances,
the fermionic and gauge form factors take the following form:
\be
\Gamma_1^i(p^2)=\frac{S_i^2}{p^2 + m_i^2}~~~
\frac{1}{p^2}\Gamma_1^A=\frac{S^2}{p^2}+ \frac{S_A^2}{p^2+m_A^2}
\ee
$S_i$, $S_A$ and $m_i$, $m_A$ are the decay constants and masses of the fermionic and gauge resonances
respectively, for $i=R,L$. 
Also, $\Gamma_0$ take the values $1$ and $p^2 /g^2$ in the fermionic and gauge case
respectively. For the simplistic scenario, with one fermion and one gauge 
boson contributing in the loop, the expressions for $\alpha$ and $\beta$  take the form,
\be
\alpha = \frac{a S_R^2}{8\pi^2\Lambda^2}\Big[\Lambda^2_{UV}-m_R^2 \log \Big( \frac{m_R^2+\Lambda^2_{UV}}{m_R^2}\Big)\Big]~~~
\beta =\frac{b g^2}{8\pi^2\Lambda^2}\Big[\Lambda^2_{UV}S^2+\Lambda^2_{UV}S_A^2-S_A^2 m_A^2 \log \Big( \frac{m_A^2+\Lambda^2_{UV}}{m_A^2}\Big)\Big]
\label{alphabeta}
\ee
after integrating out the momentum integral upto the cutoff scale$\Lambda_{UV}$.
Here, $\Lambda_{UV}=4\pi S$, $a=2 N_c$ and $b=3(N-2)/4$ where $N_c$ is the number 
of fermion colors and $N$ is the dimension of the unbroken symmetry group.

\section{Cosmological Implications of Goldstone Quintessence}
\label{implications}
\setcounter{equation}{0}
\setcounter{figure}{0}
\setcounter{table}{0}
\setcounter{footnote}{1}
In this section, let us discuss the cosmological implication of the potential,
\be \label{potential}
V(\phi)=V_0\left[1+\alpha\cos{\left(\frac{\phi}{S}\right)} + \beta\sin^2{\left(\frac{\phi}{S}\right)} \right] \,,
\ee
which has an additive factor of unity such that for the 
limiting case: $\alpha=\beta=0$ it resort back to a constant value (corresponding to the $\Lambda$CDM case). 
Let us now discuss the cosmological consequences of the pNGB framework 
at both background and linear perturbative level.

\subsection{Background Cosmology}
Considering the universe to be described by the Robertson-Walker (RW)
line-element, which is given by,
\be
ds^2= dt^2 -a^2(t) \delta_{ij}dx^i dx^j \,,
\ee
where $t$ is the co-moving time coordinate, $a(t)$ is the scale factor, 
and $\delta_{ij}$ is the Dirac-delta function, we can express the 
pNGB's energy density $\rho_\phi$ and pressure $p_\phi$ as follows:
\be
\rho_{\phi}=\frac{1}{2}\dot{\phi}^2+V(\phi)~;~~~~~~~p_{\phi}=\frac{1}{2}\dot{\phi}^2-V(\phi) \,,
\ee
where $\cdot \equiv d/dt$. Thus, scalar field's equation of state parameter $w_\phi$ is given as
\be
w_{\phi}:= \frac{p_\phi}{\rho_\phi}=\frac{\dot{\phi}^2-2V(\phi)}{\dot{\phi}^2+2V(\phi)}.
\ee
In the spatially-flat universe (for curvature $K=0$), the Friedmann and the Klein-Gordon (KG) equation are respectively given as
\bea 
&&H(t)^2 =\left(\frac{\dot{a}}{a}\right)^2 = 8 \pi G_N\left(\rho_\phi + \rho_m \right) \,, \label{friedmann}\\ 
&&\ddot{\phi}+ 3H\dot{\phi}+\frac{dV}{d\phi}=0. \,, \label{KG}
\eea
where $H(t)$ is Hubble parameter, and $\rho_m$ is the matter energy 
density. The KG equation ensures an independent conservation of scalar field and matter energy-momentum tensor which gives $\rho_m\sim a^{-3}$.
Also, the dimensionless density parameters are defined as
\be
\Omega_m:=\frac{\rho_m}{\rho_m+\rho_\phi}=\frac{8\pi G_N \rho_m}{3H^2} 
\,, \quad \Omega_\phi:=\frac{\rho_\phi}{\rho_m+\rho_\phi}=
\frac{8\pi G_N \rho_\phi}{3H^2} \,.
\ee
In the next subsection, we first describe the evolution of 
cosmological perturbations for a general non-interacting system and 
then utilise it for our composite pNGB potential (\ref{potential}). 


\subsection{Cosmological Perturbations}
At the linear perturbative level, the line-element for the 
gauge-invariant metric perturbations, also known as the Newtonian gauge, 
can be expressed as \cite{AT-book}
\be
ds^2 = e^{2N}\left[-(1+2\psi) \mathcal{H}^{-2} dN^2 + 
(1+2\Phi) \delta_{ij} dx^i dx^j \right]
\ee
where $\mathcal{H}=aH$ is the conformal Hubble parameter, $N=\log(a)$ 
denotes the number of e-foldings, and $\psi$ and $\Phi$ are the Bardeen potentials. In our analyses, we assume $\psi=-\Phi$, to nullify the effect of cosmological anistropic stress.

The evolution of linear matter density perturbations $\delta^{(m)}:= \frac{\delta \rho^{(m)}}{\rho^{(m)}}$ and field perturbations $\delta \phi$ are expressed as \cite{AT-book,PG-grow}
\begin{eqnarray}
\frac{d^2 \delta^{(m)}}{d N^2} + \frac{1}{2} \left( 1-\frac{d (\log \Omega_m)}{dN} \right) \frac{d \delta^{(m)}}{dN} = \frac{3}{2} \Omega_m \delta^{(m)}\, , \label{dmt} \\
\frac{d^2 (\delta \phi)}{dN^2} + \left(2+\frac{\mathcal{H}}{\mathcal{H} dN} \right) \frac{d (\delta \phi)}{dN} + \left( \frac{k}{\mathcal{H}} \right)^2 
-4 \frac{d \phi}{dN}\frac{d \Phi}{dN} = 2 \frac{d \bar{V}}{d \phi} \Phi \,.  
\label{dphi}
\end{eqnarray}
where $k$ denotes the scale of perturbations and $\bar{V}= V_{\text{eff}}/H^2$. 
Note that for the large-scale structure formation the observationally 
relevant scale is
$0.01h Mpc^{-1} \simeq k \simeq 0.2 h Mpc^{-1}$ which is much less than the size of the horizon and lies well inside it. Since the growth of structures are mostly dominated by the matter density perturbations, 
as the field perturbations remains highly suppressed, one can 
only explain the growth of large-scale structure formation with the matter density perturbations given by Eq.\,(\ref{dmt}).

For convenience, one can express second-order differential Eq.\,(\ref{dmt}) to a first-order differential equation as \cite{PG-grow}
\be \label{gf}
\frac{df}{d N} + f^2 + \frac{1}{2}\left(1-\frac{d (\log \Omega_m)}{dN} \right)f = \frac{3}{2}\,\Omega_m \, \quad \mbox{where} \quad 
f:=\frac{d \log(\delta^{(m)})}{dN} \,.
\ee
where $f$ is known as the growth factor. It is also convenient to 
work with a well established growth factor ansatz, given by
\be \label{ansatz}
f=\left(\Omega_m\right)^\gamma \,, \quad \mbox{where} \quad \gamma = \mbox{ growth index}
\ee
The incorporation of the growth index makes it convenient to directly observe the deviation between two models at the linear perturbative level. This is due to the fact that for the $\Lambda$CDM case, one gets $\gamma=0.555$, therefore any deviation from this value can be 
considered as a deviation with the $\Lambda$CDM value.

\section{Observational Constraints on pNGB Model}
\label{dataa}
\subsection{Data from Observations}

For the estimations of a set of parameters $p \in \{\alpha,\beta,\Omega_m,H_0\}$, we utilise a set of $H(z)$, Pantheon and BAO observations, whereas for the RSD (Redshift-space distortions) data, we have two more 
parameter $\gamma$ and $\sigma_8^0$ which is the root-mean-square 
amplitude of the fluctuations in the radius of $8Mph^{-1}$.

The $H(z)$ observations are given in Table 2 of refs. \cite{mohit} 
which consists of thirty one data points in the redshift range $z\in[0.07,1.965]$. For the BAO data, we use nine data points in the 
range $z\in[0.1,2.334]$. Note that for the scale of BAO observations 
we take the following expression for the sound horizon $r_s$ 
at the radiation drag epoch:
\be
r_s \approx \frac{44.5}{\sqrt{10 (\Omega^0_b h^2)^{3/4}+1}}\log \left(\frac{9.83}{ \Omega^0_m h^2}\right) \text{Mpc}
\ee
where the baryon fractional density at present epoch is given by $\Omega^0_b h^2=0.022$. 

For the SN1a Pantheon data we take the binned sample compiled in 
\cite{pantheon} which is a set of forty data points in the range 
$z\in[0.014,1.6123]$. For the SN1a data, the apparent magnitude 
is given by
\be
m_B=\mathcal{M}+5\log \mathcal{D}_L \,,
\ee
such that the $H$-independent luminosity distance $\mathcal{D}_L$ 
is expressed as
\be
\mathcal{D}_L =(1+z_{hel}) \int^{z}_{0} \frac{d\tilde{z}}{H(\tilde{z})/H_0} \,,
\ee
where $z_{hel}$ denotes the heliocentric frame redshift and 
$\mathcal{M}$ is the combinations of two nuisance parameters: $H_0$ 
and $M_B$ (absolute magnitude of SN1a).

\begin{figure}[htb!]
\begin{center}
\includegraphics[width=0.99\textwidth]{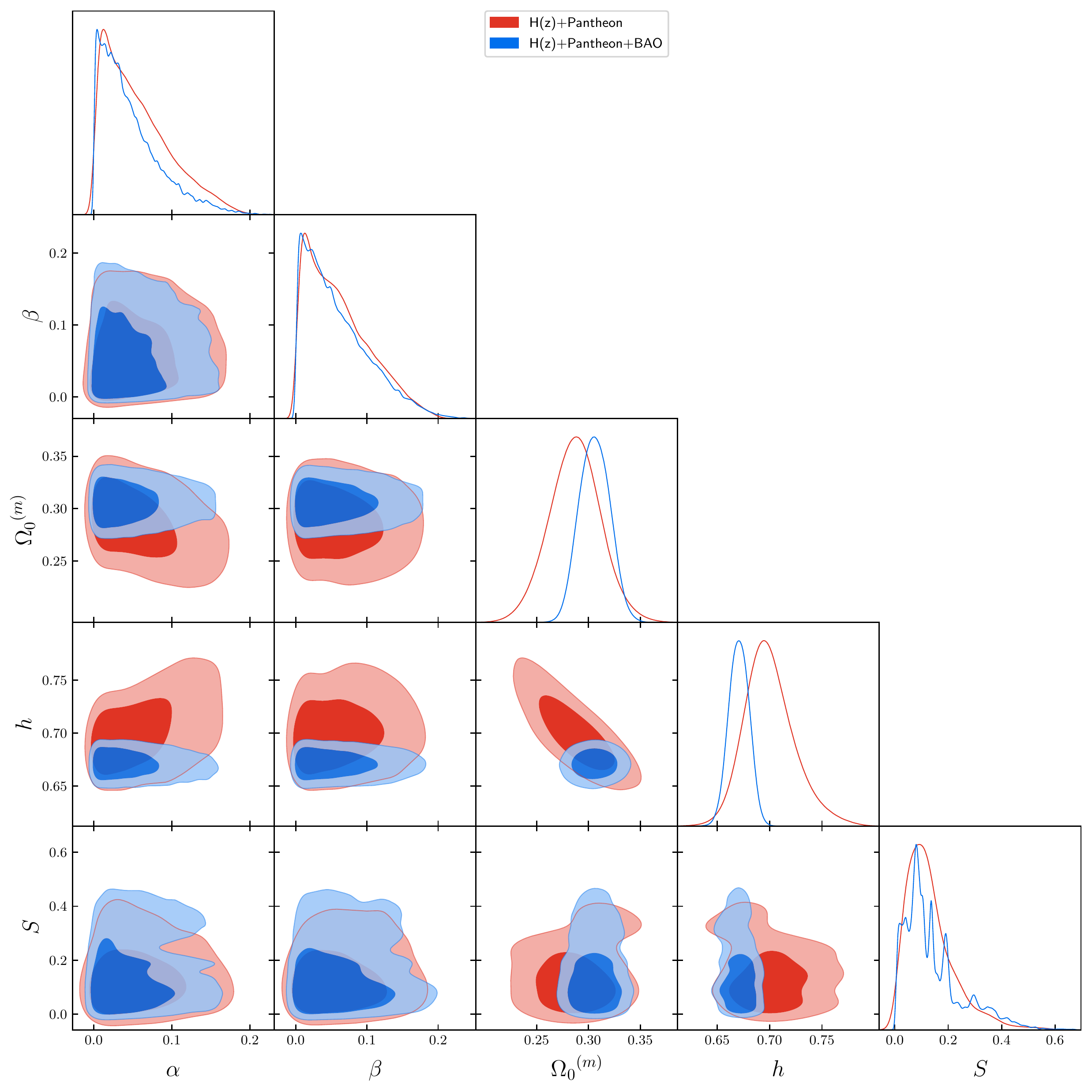}
\caption{2D contours of the combined OHD+ Pantheon(Red) and combined  OHD+ Pantheon+ BAO (Blue)constraints for
the pNGB model with chosen Goldstone potential. The individual marginalised posterior probability distributions of each parameter are also shown.}
\label{tri_1}
\end{center}
\end{figure}

\begin{figure}[htb!]
\begin{center}
\includegraphics[width=0.99\textwidth]{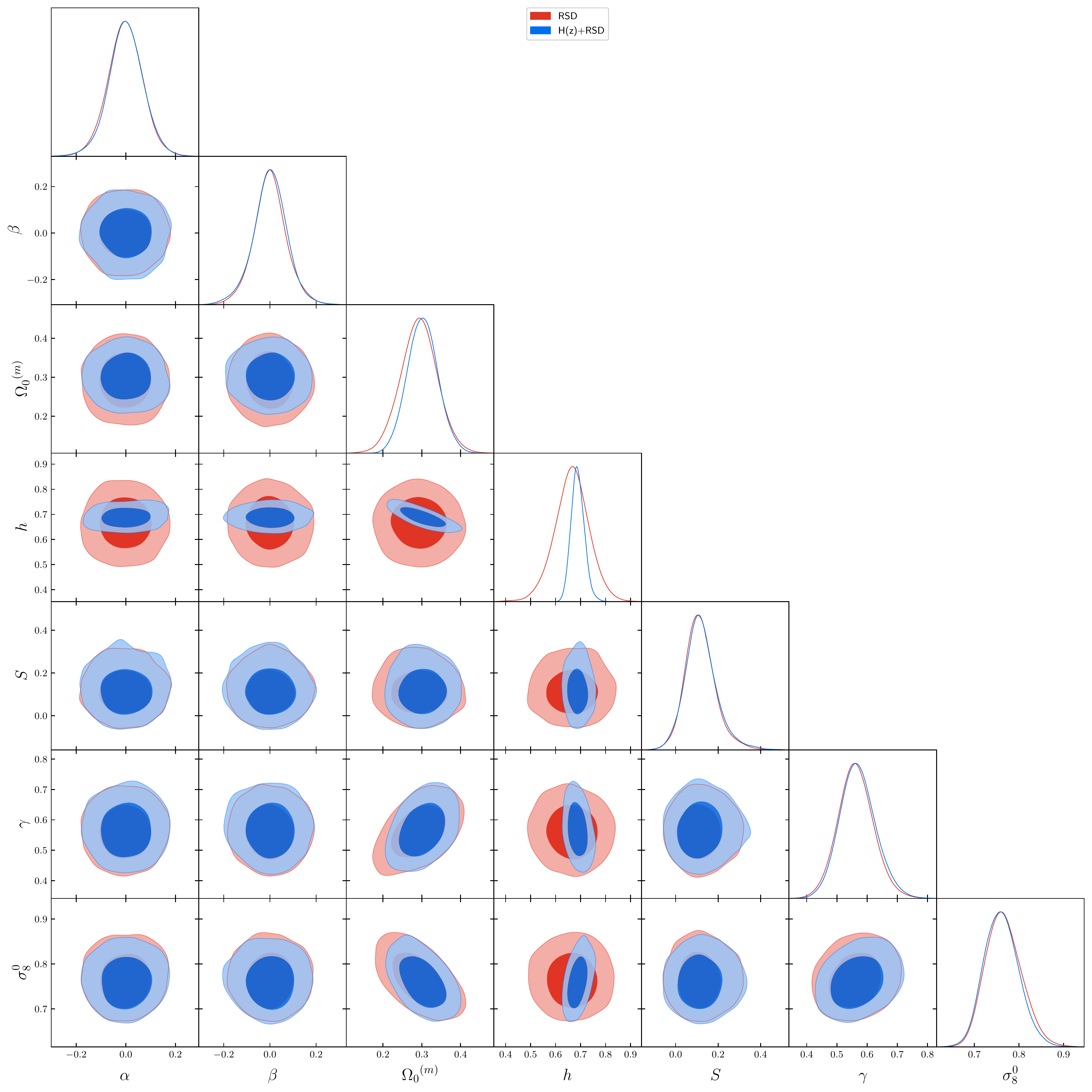}
\caption{2-D contours upto $2\sigma$ level for RSD and its 
combination with $H(z)$ data for the pNGB model with chosen 
Goldstone potential.}
\label{tri_2}
\end{center}
\end{figure}

Finally, for the RSD data, we take GOLD updated sample which consists of twenty two data points in the redshift 
range $z\in[0.02,1.944]$. The observed quantity is a combination of 
$f(z)\sigma_8(z)$ which can be written as
\be
(f\sigma_8)(z)=f(z)\sigma_8(z)= f(z) \sigma_8^0 \left(\frac{\delta^{(m)}(z)}{\delta^{(m)}(0)} \right) \,.
\ee
In the next subsection we will implement the statistical technique 
for our parametric estimations.

\subsection{Statistical Analysis}
We use the Metropolis Algorithm of Markov Chain Monte Carlo (MCMC) 
technique for our parametric estimations on the basis of the 
Likelihood maximization such that Likelihood function $\mathcal{L}$ 
is defined as
\be
\mathcal{L} = e^{-\chi^2 /2} \,.
\ee
For $H(z)$ data, $\chi^2$ is given by
\be
\chi^2_H= \sum_i\left(\frac{H_{th}(p,z)-H_{obs}(z)}{\sigma_H} \right)^2
\ee
where $H_{th}$ is the theoretically calculated Hubble parameter, 
$H_{obs}(z)$ is the corresponding observed value at the redshift $z$, 
and $\sigma_H$ is the uncertainty in the measurements. For the 
Pantheon data, the $\chi^2$ can be written as
\be
\chi^2_{SN} = \Delta m_B^T \cdot C^{-1}\cdot \Delta m_B \,,
\ee
where $\Delta m_B=m_{obs}-m_{th}$. Note that the covariance matrix 
$C$ is given by 
\be
C= C_{sys} + D_{stat} \,,
\ee
where $C_{sys}$ is the systematic uncertainty and $D_{stat}$ is 
the statistical uncertainty in measurements. 

For the RSD data, the $\chi^2$ is given by
\be
\chi^2_{RSD} =  \Delta A^T \cdot C_{wiggleZ}^{-1}\cdot \Delta A \,,
\ee
where $\Delta A\equiv (f\sigma_8)|_{th}(z)-(f\sigma_8)|_{obs}(z)$, 
and $C_{wiggleZ}$ is the covariance matrix between three WiggleZ 
data points.

In order to constrain the parameters, we use the Markov Chain 
Monte Carlo (MCMC) method to obtain the median best-fit values. 
The priors for that are given as follows:
\bea
-0.3\leq \!\!\!\! &&\alpha \leq 0.3 \,, \quad -0.3\leq \beta \leq 0.3 \,, 
\quad 0.1 \leq \Omega_0^{(m)}<0.5 \,, \quad 0.55\leq h \leq 0.85 \,, \nonumber \\
\quad && 0 < f < 1 \,,  \quad 0.1\leq \gamma \leq 0.8 \,, \quad 
0.65 \leq \sigma_8^{(0)} \leq 0.9 \,.
\eea

\begin{table*}[ht]
\centering
\renewcommand{\arraystretch}{1.7}
\resizebox{\textwidth}{!}{
\begin{tabular}{||c|c|c|c|c|c|c|c||}
\hline
 & \multicolumn{7}{c||}{\small Parametric estimations} \\
 &  \multicolumn{7}{c||}{\footnotesize (best fit \& $1\sigma$ limits)} \\
\cline{2-8}{}
%
 & $\alpha$ & $\beta$ & $\Omega_0^{(m)}$  & $h$ & $S$ & $\gamma$ & 
 $\sigma_8^0$ \\
\hline\hline
$H(z)$+Pantheon & $ 0.046^{+0.053}_{-0.033}$ & $0.052^{+0.057}_{-0.037}$  & $0.287^{+0.024}_{-0.025}$ & $0.697^{+0.025}_{-0.021}$  &  $0.112^{+0.103}_{-0.068}$  & - & -\\
\hline
$H(z)$+Pantheon+BAO & $ 0.036^{+0.048}_{-0.026}$ & $0.048^{+0.058}_{-0.034}$  & $0.306^{+0.015}_{-0.015}$ & $0.671^{+0.01}_{-0.01}$ & $0.108^{+0.148}_{-0.065}$ & - & - \\
 \hline
RSD & $<0.069$ & $<0.067$  & $0.293^{+0.046}_{-0.047}$ & $0.666^{+0.065}_{-0.068}$ & $0.111^{+0.071}_{-0.063}$ & $0.561^{+0.059}_{-0.056}$ &  
$0.763^{+0.042}_{-0.038}$ \\
\hline
$H(z)$+RSD & $<0.068$ & $<0.068$  & $0.302^{+0.038}_{-0.039}$ & $0.687^{+0.028}_{-0.025}$ & $0.114^{+0.076}_{-0.065}$ & $0.567^{+0.061}_{-0.056}$ &  
$0.759^{+0.039}_{-0.037}$ \\
\hline\hline
\end{tabular}
}
\caption{\footnotesize Best fit values of parameters with their $1\sigma$ confidence limits obtained for background level and perturbation dataset,  as well as for the combination of them.}
{\label{E-tab1}}
\end{table*} 
\begin{figure}[htb!] 
\begin{center}
\includegraphics[width=0.6\textwidth]{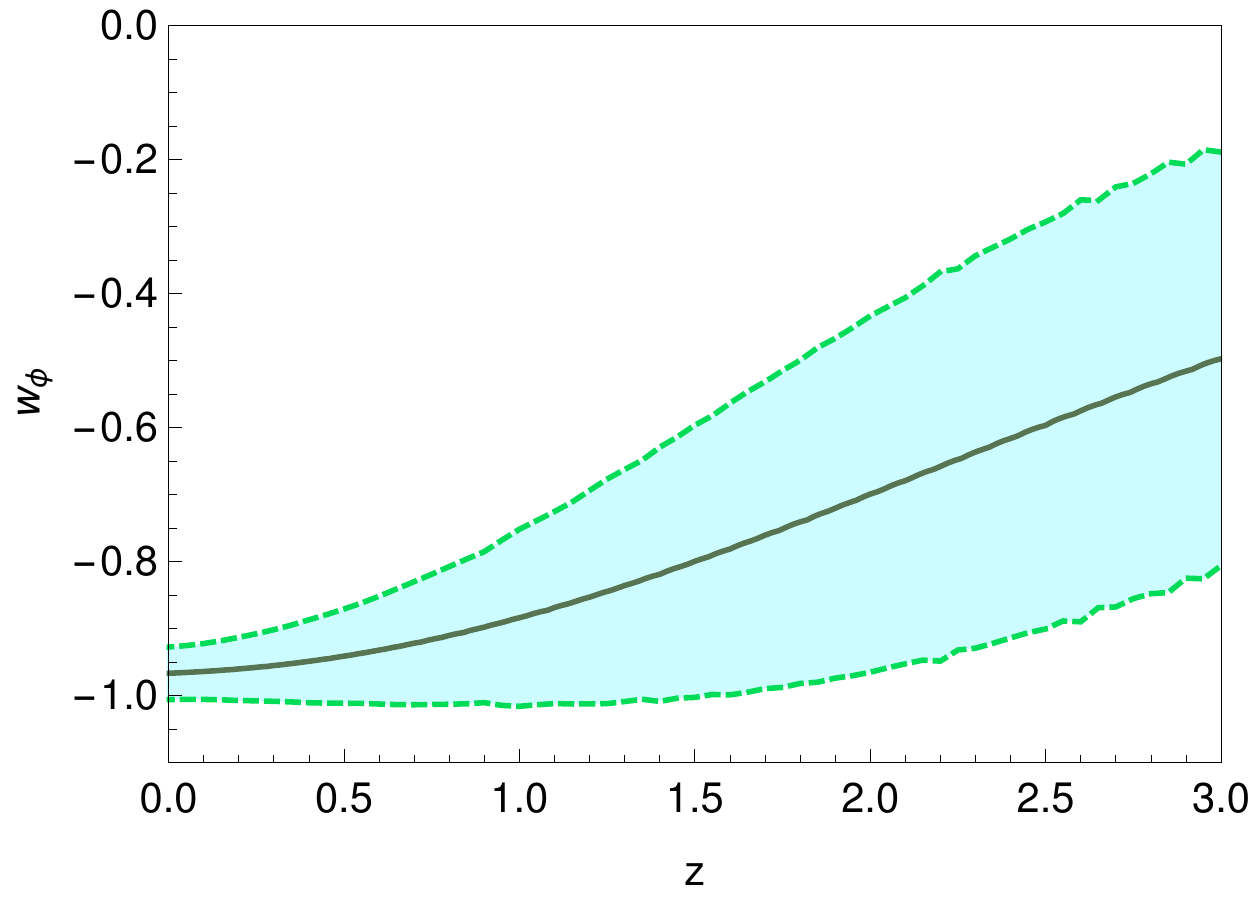} 
\end{center}
\caption{Upto $1\sigma$ evolutionary profile of pNGB quintessence equation of state 
$w_\phi$ with $z \in [0,3]$.}
\label{fig:wde}
\end{figure}

The obtained values for the constrained parameters are shown in table\,
(\ref{E-tab1}) and the allowed parametric region between the parameters 
are for background and perturbations are shown in fig.\,(\ref{tri_1}) and (\ref{tri_2}), respectively. In the latter, we see that $h$ is more tightly constraint for the combination of $H(z)$ and RSD than for the $H(z)$ alone.  Note that we have obtained sub-Planckian values of $S$ in all cases. The evolution of 
$w_\phi$ with $z$ for one such case i.e. for $H(z)$+Pantheon best-fits 
are shown in fig.\,(\ref{fig:wde}). In other two cases also, $w_\phi$ behaves accordingly. In that figure, one finds that 
although near $z \to 0$, the model behaves as $\Lambda$CDM but at high 
red shifts there is small oscillations in $w_\phi$. This is due to the 
fact that the potential $V(\phi)$, for non-zero $\alpha$ and $\beta$, 
is oscillating in nature. For small $z$, these oscillations are highly 
suppressed due to the accelerated expansion of the universe.


\subsection{Connection with UV Physics and Resonances}
\begin{figure}[htb!]
\begin{center}
\includegraphics[width=0.4\textwidth]{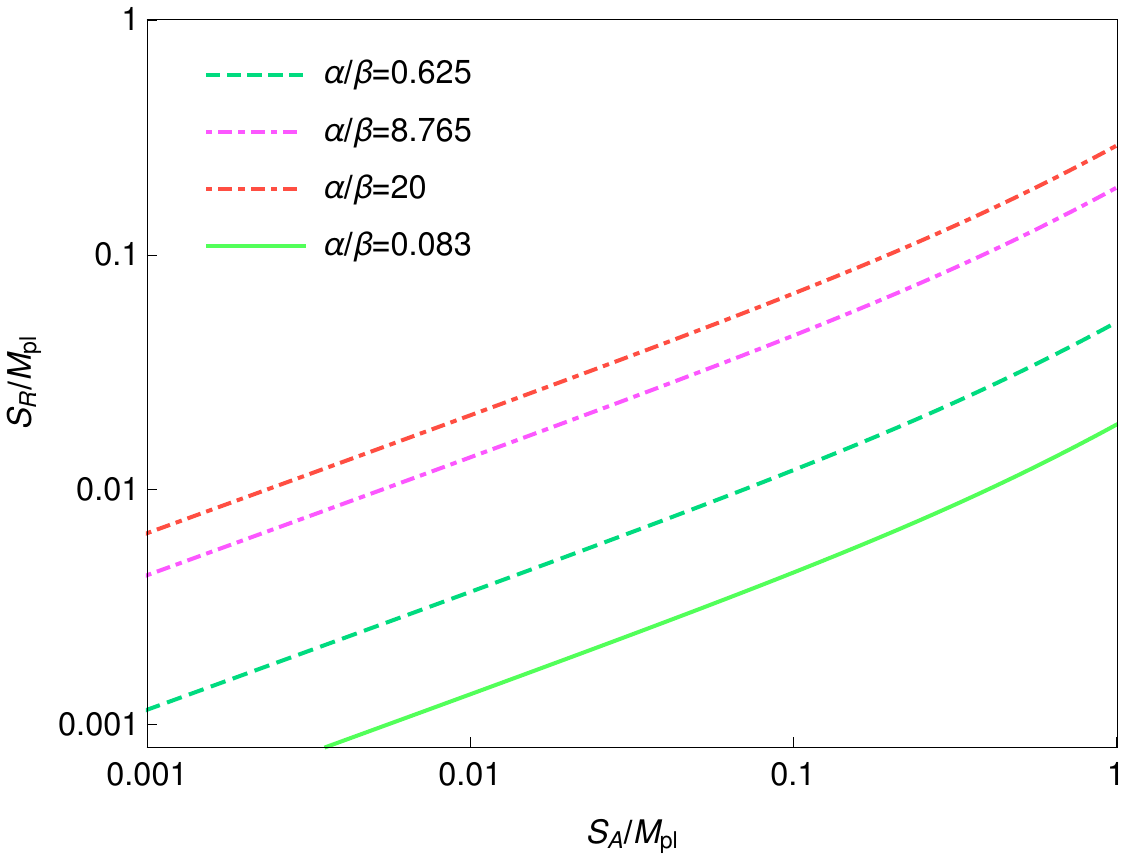}
\includegraphics[width=0.4\textwidth]{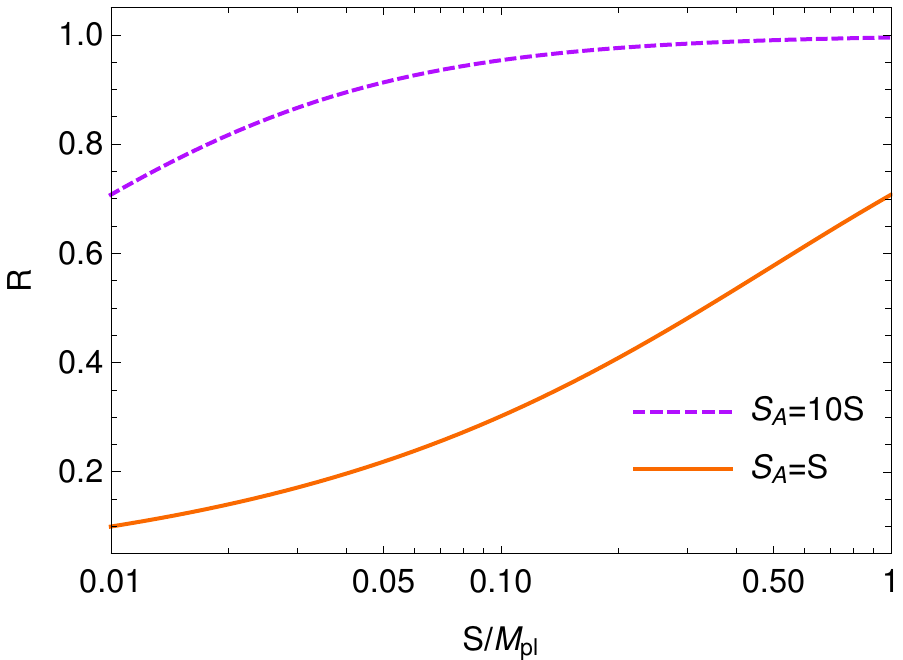}
\caption{(left) Variation of $S_R$ is shown with $S_A$ with the ratio $\alpha/\beta$, for the two best fit values 
with (Dashed) and without (Solid) matter perturbation, and two arbitrary values of $\alpha/\beta$ (Dotted Dashed). 
$S_A$ is assumed to be at scale $S$. 
(right) Variation of the ratio as defined in Eq:~\ref{reso} with the symmetry breaking scale, for two cases, $S_A=S$ (Solid Red)and $S_A=10S$(Dashed Magenta). }
\label{resonance}
\end{center}
\end{figure}

By using the expression of $\alpha$ and $\beta$ in Eq.~(\ref{alphabeta}), 
we obtain the following relation by demanding that the  
cutoff dependence of the quadratic terms cancel exactly
\bea
\frac{\alpha}{\beta}=\frac{b g^2 S_A^2}{a S_R^2};~~
R=\frac{S_A^2}{S^2+S_A^2} = \frac{m_R^2\log \Big( \frac{m_R^2+\Lambda^2_{UV}}{m_R^2}\Big)}{m_A^2\log \Big( \frac{m_A^2+\Lambda^2_{UV}}
{m_A^2}\Big)} \,.
\label{reso}
\eea
In fig.~\ref{resonance}(left) we show the variation of the 
fermionic and bosonic decay constants for the two best-fit scenarios, 
without ($\alpha/\beta$=0.625) and with ($\alpha/\beta$=0.083)
matter perturbation. We have seen before in fig.\,\ref{tri_2} that 
when we take the matter perturbation into account, 
the distribution of $\alpha$ and $\beta$ is more symmetric around origin, 
and in general prefers smaller values compared to the 
case with no matter perturbation fig.\,\ref{tri_1}. In both the cases we obtain $S_A > S_R$. 
We have also checked two random scenarios, when $\alpha/\beta$ is larger, 
and that gives us $S_R > S_A$. We also show the variation in the ratio 
$R$ in fig.\,\ref{resonance}(right) with the scale $S$. 
In this plot we get $R < 1$ in most of the region, implying 
a degeneracy between fermionic and bosonic states given by $m_A> m_R$.
We show that if $S_A$ moves closer to $S$, $R\rightarrow 1$, 
which in turn implies that the degeneracy is removed. In our case, $S_A$ can 
take the values only upto $10 S$, for physics below the Plank scale as value of $S$
ranges from 0 to 0.9, as shown in the fit result. 
Even though it is straightforward to check from the analytical expression that 
this condition will be satisfied for any scale factor between $\alpha$ and $\beta$, here we confirm this scenario with the current data as confronted by the composite pNGB quintessence. 
 
\section{Conclusions and Discussions}
\label{discussions}
\setcounter{equation}{0}
\setcounter{figure}{0}
\setcounter{table}{0}
\setcounter{footnote}{1}

In this paper we have explored the possibility of having a late-time 
cosmic acceleration due to a pNGB field, which comes out of a 
spontaneous symmetry breaking in the early universe and develops a CW potential at loop level. Motivated from the Composite Higgs like scenario, here 
the potential has two oscillatory terms, unlike the standard pNGB potential that was proposed before to solve the DE dynamics. In the model under consideration, it is possible to obtain a sub-Planckian value of the pNGB decay constant $S$ while satisfying all observational data.  
In this study we have found that the tuning between the coefficients of these two terms ($\alpha$, $\beta$) are the key ingredient to achieve the evolution of the equation of state of field $w_\phi$ in the right range while keeping $S$ sub-Planckian. Which means that this DE model is perfectly within the limits of the effective field theory. Moreover, the parameter dependence of this potential is not ad-hoc, but comes from the integrated out dynamics above scale $S$. 
 
We have shown that the pNGB potential \ref{potential} in the limit 
$\alpha,\beta \to 0$ approaches to that of the $\Lambda$CDM at which, 
as expected, it gives back the $\Lambda$CDM consistent results.
In order to check the viability of the pNGB motivated quintessence 
scenario, we have discussed its impact on both cosmological and linear 
perturbative level in presence of the dust-like matter. We have found 
that for the $H(z)$+Pantheon dataset the best-fit value of the Hubble constant comes out to be slightly higher due to the non-zero best-fit 
values of model parameters $\alpha$ and $\beta$ which slightly 
helps in reducing the current Hubble tension problem 
(see Table\,\ref{E-tab1}). For the RSD and $H(z)$+RSD dataset, 
the best-fit of growth index comes out to be slightly larger than the 
$\Lambda$CDM value i.e. 0.555.
In a nutshell, the proposal of having a pNGB like quintessence 
scenario, which as a potential like Composite Higgs Model, satisfy the late-time observational constraints and shows small deviations from the $\Lambda$CDM model. However, 
checking the conditions for the dynamical stability of the above 
scenario which consist of trigonometric functions in the potential is 
crucial. We are aiming to perform that in our next work which we will 
try to report soon.

\textbf {Acknowledgments:} We thank M. Sami for the interesting discussions during the work. The work of MRG is supported by the Department of Science and Technology, Government of India under the Grant Agreement number IF18-PH-228 (INSPIRE Faculty Award). The work of NK is supported by the Department of Science and Technology, Government of India under the SRG grant, Grant Agreement Number SRG/2022/000363. The work of MRG and NK is also supported by Science and
Engineering Research Board(SERB), Department of Science and Technology(DST), Government of
India under the Grant Agreement number CRG/2022/004120(Core Research Grant).


\end{document}